\documentclass[11pt,a4paper]{article}

\usepackage{amsmath,amssymb,amsthm}
\usepackage{graphicx}
\usepackage{bm}
\usepackage{natbib}
\usepackage{url}
\usepackage{authblk}
\usepackage{geometry}
\usepackage{setspace}
\usepackage{mathtools} 
\usepackage{subcaption}
\title{Numerical optimization for the compatibility constant of the lasso}

\author{Kei Hirose}
\affil{Institute of Mathematics for Industry, Kyushu University, Japan}
\date{}

\newtheorem{proposition}{Proposition}

\theoremstyle{definition}
\newtheorem{definition}{Definition}
\theoremstyle{remark}
\newtheorem{remark}{Remark}

\begin{document}
\maketitle

\begin{abstract}
The compatibility constant plays an important role in evaluating the prediction error of the lasso in high-dimensional settings. However, the computation of the compatibility constant is generally difficult because it is a complicated nonconvex optimization problem. In this study, we present a numerical approach to compute the compatibility constant when the support of true regression coefficients is given. We show that the optimization problem reduces to a quadratic programming (QP) once the signs of the nonzero coefficients are specified. In this case, the compatibility constant can be obtained by solving QPs for all possible sign combinations. We also formulate a mixed-integer QP (MIQP) approach that can be applied when the number of true nonzero coefficients is large. We investigate the finite-sample behavior of the compatibility constant for simulated data under various parameter settings and compare the prediction error with its theoretical upper bound. The behavior of the compatibility constant in finite samples is also investigated through a real data analysis.
\end{abstract}


\section{Introduction}
The success of the lasso (least absolute shrinkage and selection operator; \citealp{tibshirani1996lasso}) lies in the computational efficiency \citep{efron2004least,friedman2010regularization,boyd2011distributed,boyd2004convex,Stellato2020OSQP}
and strong theoretical foundations in high-dimensional settings \citep{bickel2009simultaneous,van2009conditions,buhlmann2011statistics,zhang2010nearly,zhao2006model,wainwright2009sharp,meinshausen2009lasso}. In particular, the theory of the lasso has been shown to provide good estimation and prediction accuracy even when the number of variables exceeds the number of observations.  For example, some regularity conditions, such as the compatibility condition, restricted eigenvalue condition, and irrepresentable condition, are sufficient conditions for obtaining a desirable property, such as oracle inequality and model selection consistency \citep{zhao2006model,bickel2009simultaneous,van2009conditions,buhlmann2011statistics}.  The theory of the lasso shares many similarities with that of compressive sensing, where the restricted isometry property has been shown to guarantee sparse recovery \citep{candes2005decoding,donoho2006compressed}.

This paper focuses on the compatibility condition \citep{buhlmann2011statistics}, one of the crucial conditions for establishing the oracle inequality of the lasso estimator.  The compatibility condition is weaker than many other assumptions,
including the restricted eigenvalue, irrepresentable, and restricted isometry conditions
\citep{van2009conditions,bickel2009simultaneous,buhlmann2011statistics}.  Therefore, it is crucial to investigate whether the compatibility condition holds; if it does not, there is a possibility that the lasso can fail to achieve good prediction accuracy.

The compatibility condition provides an upper bound on prediction error in a high-dimensional setting.  This upper bound is highly dependent on a quantity referred to as the compatibility constant.  As the compatibility constant increases, the theoretical upper bound on the prediction error decreases \citep{buhlmann2011statistics}.   Therefore, the computation of the compatibility constant is essential for understanding the finite-sample behavior of the upper bound on the prediction error.  Once we know how to compute the compatibility constant, we can evaluate not only the upper bound on the prediction error but also how far the upper bound deviates from the actual prediction error. In addition, we can study how the compatibility constant changes with the number of observations and the dimensionality. Such empirical investigations would provide a valuable step toward bridging the gap between lasso theory and its practical performance.

Nevertheless, the compatibility constant has rarely been computed for a given design matrix. This is partly because most theoretical research has focused on establishing upper bounds for evaluating the prediction error, and the compatibility constant is typically used only to provide a sufficient condition for guaranteeing such upper bounds. In other words, the compatibility constant is often treated as a theoretical quantity rather than being evaluated numerically for specific design matrices. Another reason is the intrinsic difficulty in computing the compatibility constant.  First, the computation of the compatibility constant is generally impossible, as it depends on the support of the true regression coefficients (i.e., the set of indices corresponding to nonzero coefficients), which cannot be known in advance. Even if we know the support in advance, the computation is still challenging in practice. The computation is similar to finding the smallest eigenvalue of the sample covariance matrix but can differ and may be significantly larger than this eigenvalue.  This is because the associated vector (though not exactly an eigenvector) must lie within a nonconvex set defined by the support of the true nonzero coefficients \citep{buhlmann2011statistics}.  As a result, an analytical approach is generally infeasible. Although a numerical approach is helpful, the problem reduces to a complicated nonconvex optimization problem.

In this study, we present a numerical approach to compute the compatibility constant when the support of true regression coefficients is known. We show that when the signs of the nonzero coefficients are provided, the computation reduces to a Quadratic Programming (QP). Using this property, the compatibility constant can be obtained by solving QPs for all possible sign combinations and selecting the smallest value. When the number of true nonzero coefficients is large, it is difficult to obtain the compatibility constant with the above QP-based procedure due to a huge number of combinations of the signs of the true nonzero regression coefficients. To handle the combinatorial complexity, we formulate the problem as a mixed-integer quadratic programming (MIQP, \citealp{BertsimasTsitsiklis1997,BertsimasWeismantel2005}) incorporating Big-$M$ constraints \citep{NemhauserWolsey1988} or special ordered set type 1 (SOS1) variables \citep{BealeTomlin1970,Williams2013}.

We investigate the numerical behavior of the compatibility constant for simulated data under a wide variety of parameter settings.  The results show that the compatibility constant is highly dependent on the number of observations; therefore, we must take it into account when investigating the prediction error. We also investigate how the bound on the prediction error changes at each step of the proof of the oracle inequality. Furthermore, we provide a heuristic approach to estimate the compatibility constant when the support of true regression coefficients is unknown. This approach is applied to the analysis of the S\&P500 dataset to demonstrate the relevance to the simulation results.

The remainder of this paper is organized as follows. In Section 2, we briefly review the compatibility condition and its related topics.  In Section 3, we provide a method for computing the compatibility constant based on QP and MIQP.  Section 4 presents numerical simulations to compute the compatibility constant under a wide variety of settings, and compares the prediction error and its upper bound constructed with the compatibility constant.  In Section 5, we provide an actual data analysis to illustrate the behavior of the estimate of the compatibility constant.  Section 6 concludes our work.

\section{Preliminaries}
Let $\bm y = (y_{1},\dots,y_{n})^\top \in \mathbb{R}^n$ be a response vector and $X \in \mathbb{R}^{n\times p}$ be a design matrix. Suppose that predictors are scaled and responses are centered; that is, without loss of generality, the intercept is 0 and ${\rm diag}(X^\top X)=nI_p$, where $I_p$ is a $p \times p$ identity matrix. Consider the linear regression model
\begin{equation}
    \bm y = X \bm \beta + \bm \varepsilon, \label{eq:regmodel}
\end{equation}
where $\bm \beta$ is a regression coefficient vector and $\bm \varepsilon$ is an error vector. Consider the lasso problem,
$$\frac{1}{2n}\|\bm y - X\bm \beta\|_2^2 + \lambda \|\bm \beta\|_1,$$
where $\lambda>0$ is a regularization parameter which controls the fit to the data and the degree of sparsity.

Let $S := \{ j \mid \beta_j \neq 0\}$, and let $s := |S|$ denote its cardinality. For constructing the theory of the lasso for high-dimensional data, we assume that $s$ is sufficiently small compared with $n$ and $p$; in other words, true regression coefficients are sufficiently sparse.  We may also consider the case $n < p$, but this is not necessary; we assume that $\frac{\log p}{ n}\rightarrow 0$ as $n \rightarrow \infty$.

The compatibility condition is one of the most important assumptions for deriving an upper bound on the prediction error in high-dimensional settings. Let 
$\hat{\Sigma} = X^\top X / n$ be the Gram matrix of the design matrix.
The compatibility condition, a condition for the design matrix $X$, is defined as follows:
\begin{definition}
    The compatibility condition (\citealp{buhlmann2011statistics}, p.~106) is met if there exists some $\phi>0$ such that, for any $\bm v \in \mathbb{R}^p$ satisfying $\|\bm v_{S^c}\|_1 \leq 3\|\bm v_{S}\|_1$, it holds that
    \begin{equation}
        \|\bm v_{S}\|_1^2 \leq  s\frac{\bm v ^\top \hat{\Sigma}\bm v}{\phi^2} \label{eq:compatibility}
    \end{equation}
    Here, $\phi>0$ is referred to as the compatibility constant.
\end{definition}
The compatibility constant is related to the smallest eigenvalue of $\hat{\Sigma}$ because $\|\bm v_{S}\|_1^2 \leq s\|\bm v_{S}\|_2^2$.  However, $\phi$ is generally larger than this eigenvalue because of the constraint $\|\bm v_{S^c}\|_1 \le 3\|\bm v_S\|_1$ \citep{buhlmann2011statistics}.  The constraint $\|\bm v_{S^c}\|_1 \le 3\|\bm v_S\|_1$ restricts the vectors to be in a nonconvex set where the main direction of the vector mostly aligns with $S$. In other words, feasible directions cannot deviate too far from the coordinates indexed by $S$. 

To illustrate this intuition, consider the compound-symmetry structure, $\hat{\Sigma} = \rho \bm{1}_p\bm{1}_p^\top + (1-\rho)I_p$ with $\rho$ being the correlation coefficient between any two predictors.  When $\rho>0$, the largest eigenvalue of $\hat{\Sigma}$ is $1+(p-1)\rho$ and the corresponding eigenvector is $\bm{1}_p$. The other eigenvalues are all $1-\rho$ and the corresponding eigenvectors are those orthogonal to $\bm{1}_p$. However, a direction $\bm{1}_p$ cannot satisfy the constraint $\|\bm v_{S^c}\|_1 \le 3\|\bm v_S\|_1$ when $|S^c| \gg |S|$.  Consequently, the feasible region excludes the direction associated with the largest eigenvalue, and any feasible $\bm v$ must deviate substantially from the direction $\bm{1}_p$.  Therefore, when $\rho \approx 1$, $\bm v^\top \hat\Sigma \bm v$ that satisfies $\|\bm v_{S^c}\|_1 \le 3\|\bm v_S\|_1$ must be small, leading to a small compatibility constant.  In contrast, when $\rho \approx 0$, the quadratic form $\bm v^\top \hat\Sigma \bm v$ does not vary significantly depending on the direction of $\bm v$.  As a result, the compatibility constant tends to be larger.

Under the compatibility condition, we obtain the following result (\citealp{Tibshirani2017Sparsity}).
\begin{proposition}
Suppose that $\lambda$ satisfies $\lambda \ge 2\sigma \sqrt{2/n (1+\log \left( \frac{p}{\delta} \right))}$ with $\delta \in (0,1)$. Then, with probability at least $1-\delta$, we have
\begin{equation}
\frac{1}{n} \| X\hat{\bm\beta} - X\bm\beta \|_2^2 \leq \frac{9 s \lambda^2}{\phi^2}. \label{eq:error}    
\end{equation}
In particular, plugging $\lambda = 2\sigma \sqrt{2/n (1+\log \left( \frac{p}{\delta} \right))}$ into the above equation, we have 
\begin{equation}
\frac{1}{n} \| X\hat{\bm\beta} - X\bm\beta \|_2^2 \leq \frac{72\sigma^2 s (1+\log(p / \delta))}{n \phi^2}\label{eq:error2}
\end{equation}
\label{prop:lassoaccuracy}
\end{proposition}
For details of the result and its proof, please refer to Section 4.2 of \citet{Tibshirani2017Sparsity}.  Some of the proofs are also given in Appendix \ref{app:MSPEderivation}.  We may obtain  results whose constant value is slightly different from \eqref{eq:error} (e.g., 9 might be changed to 4; \citealp{buhlmann2011statistics}), but the rate of convergence remains unchanged; roughly speaking, the error converges to 0 with high probability when $\log p /n  \rightarrow 0$ as $n \rightarrow \infty$.   

The compatibility constant $\phi$ plays an important role in evaluating the error bound.  Since the upper bound in Proposition~\ref{prop:lassoaccuracy} is proportional to $\phi^{-2}$, the magnitude of the compatibility constant $\phi$ directly determines how large the bound of the prediction error is.  When $\phi$ is large, the upper bound becomes small, which implies that the lasso may achieve a meaningful prediction accuracy if the tuning parameter is properly chosen.

Several conditions other than the compatibility condition have been proposed for establishing theoretical properties of the lasso \citep{van2009conditions}. In particular, the irrepresentable condition and restricted eigenvalue condition are two popular conditions. Their purposes differ depending on the aspect of the problem. The irrepresentable condition is closely related to model selection consistency \citep{zhao2006model}, while the restricted eigenvalue condition and the restricted isometry property are often used to analyze prediction error \citep{bickel2009simultaneous,candes2005decoding}.  On the other hand, the compatibility condition is related to oracle inequalities as described in Proposition \ref{prop:lassoaccuracy}. Among these conditions, the compatibility condition is known to be the weakest condition that still guarantees oracle inequalities. In fact, \citet{van2009conditions} show that the compatibility condition is implied by both the restricted eigenvalue condition and the restricted isometry property. For this reason, the compatibility condition can be considered a flexible assumption for studying the prediction performance of the lasso in high-dimensional settings.

\section{Numerical optimization for compatibility constant}

As the analytical derivation of the compatibility constant is not easy due to the constraint $\|\bm v_{S^c}\|_1 \leq 3\|\bm v_S\|_1$, we provide a numerical optimization method.  First, we reformulate the compatibility condition.  The compatibility constant in \eqref{eq:compatibility} can be computed with the following optimization problem:
\begin{equation}
\phi^2
=
\min_{ \{\bm v \mid \bm v \neq \bm 0 \}}
\left\{
\frac{s\bm{v}^\top\hat{\Sigma}\bm{v}}{\|\bm v_S\|_1^2} \ \middle|\; 
\|\bm{v}_{S^c}\|_1 \le 3\|\bm{v}_S\|_1
\right\}.
\label{eq:problem1}
\end{equation}
The above problem turns out to be scale-invariant. Indeed, for any nonzero scalar $a$, the vectors $\bm{v}$ and $a\bm{v}$ yield the same compatibility constant.  Therefore, without loss of generality, we may impose the constraint $\|\bm v_S\|_1=1$, in which case the constraint $\|\bm v_{S^c}\|_1 \le 3\|\bm v_S\|_1$ reduces to $\|\bm v_{S^c}\|_1 \le 3$. Hence, the problem \eqref{eq:problem1} can be reformulated as
\begin{equation}
\phi^2
=
\min_{\bm v}
\left\{
s\bm{v}^\top\hat{\Sigma}\bm{v} \ \middle|\; \|\bm{v}_S\|_1=1, \ 
\|\bm{v}_{S^c}\|_1 \le 3
\right\}.
\label{eq:problem2}
\end{equation}
The above problem appears to be a quadratic programming (QP). However, the constraint $\|\bm{v}_S\|_1=1$ is not a convex set. As a result, it would be difficult to directly apply the QP or other convex optimization tools.

\subsection{Quadratic Programming}
To address the above-mentioned issue, we consider a much simpler problem where the sign of each element of $\bm{v}_S$, say $\bm z \in \{\pm1\}^s$, is fixed. In this case, the constraint, $\|\bm{v}_S\|_1=1$ is expressed as
\begin{equation*}
    \sum_{j \in S} z_jv_j = 1
\end{equation*}
with a constraint that $z_jv_j \geq 0 \ (\forall j \in S)$. 
Since 
$$\left\{\bm v \in \mathbb{R}^p \middle|\; \sum_{j \in S} z_jv_j = 1, \  z_jv_j \geq 0 \ (\forall j \in S)\right\}$$
is a convex set, we can easily obtain $\phi^2$ for given $\bm z$, say $\phi^2_{\bm z}$, by solving the following problem: 
\begin{equation}
\begin{alignedat}{2}
\phi^2_{\bm z}:=& \min_{\bm v\in\mathbb{R}^p,\;\bm u \in\mathbb{R}^{|S^c|}}
  s\bm v^\top \hat{\Sigma} \bm v\\
\text{subject to} \quad & \sum_{j\in S}  z_j\,v_j = 1,\\
 & z_j\,v_j \geq 0 &  (\forall j\in S),\\
 &-u_j \leq v_j \leq u_j \quad  & (\forall j\in S^c),\\
 &u_j \geq 0 & (\forall j\in S^c),\\
 &\sum_{j\in S^c} u_j \le 3 .
\end{alignedat}
\label{eq:problem3}
\end{equation}
All constraints in \eqref{eq:problem3} are linear and the feasible region is convex. The objective function $\bm v^\top \hat{\Sigma} \bm v$ is also convex because $\hat{\Sigma}$ is positive semidefinite.  Hence, for a fixed sign pattern, the optimization problem \eqref{eq:problem3} becomes a standard convex quadratic program (QP). We can obtain $\phi^2_{\bm z}$ with a solver for quadratic programming, such as the OSQP library \citep{Stellato2020OSQP}.
The compatibility constant is obtained by solving \eqref{eq:problem3} for all possible sign patterns of $\bm z$ as below:
\begin{align}
    \phi^2=\min_{\,z_j\in\{\pm1\}^s}\ \phi_{\bm z}^2.\label{eq:problem4}
\end{align}

\begin{remark}
The compatibility constant is invariant under a global sign flip of $\bm v$. Indeed, if $\bm v$ satisfies the condition in \eqref{eq:problem2}, $-\bm v$ also satisfies both $\|\bm v_S\|_1=1$ and $\|\bm v_{S^c}\|_1 \le 3$, and the objective value is unchanged because $\bm v^\top \hat{\Sigma} \bm v = (-\bm v)^\top \hat{\Sigma} (-\bm v)$. Therefore, when enumerating sign patterns of $\bm v_S$, we only need to search $2^{s-1}$ combinations of signs.
\end{remark}

\subsection{Mixed-Integer Quadratic Programming}
The problem \eqref{eq:problem4} can be easily solved when the number of elements in active set, $s$, is sufficiently small.  However, since the number of sign patterns in \eqref{eq:problem4} is $2^{s-1}$ and grows exponentially, the computation of \eqref{eq:problem4} becomes infeasible when $s$ is large.

To avoid enumerating all $2^{s-1}$ sign patterns explicitly, we formulate the problem as a mixed-integer quadratic program (MIQP, \citealp{BertsimasTsitsiklis1997,BertsimasWeismantel2005}).  For each $j\in S$, we introduce nonnegative variables $v_j^+, v_j^- \ge 0$ and a binary variable $b_j\in\{0,1\}$ to express $v_j$ as follows:
\begin{equation*}
v_j = v_j^+ - v_j^-, \quad
v_j^+ \le M b_j,\quad v_j^- \le M(1-b_j),
\end{equation*}
where $M>0$ is a sufficiently large constant. This formulation is referred to as Big-$M$ formulation \citep{NemhauserWolsey1988}. With this formulation, the constraint $\|\bm{v}_S\|_1=1$ can be written as a linear equality
$$
\sum_{j\in S} (v_j^+ + v_j^-) = 1,
$$
and the problem \eqref{eq:problem2} reduces to the following mixed-integer quadratic program (MIQP):
\begin{equation}
\begin{alignedat}{2}
\min_{\substack{
\bm{v}_S^+,\bm{v}_S^- \ge 0,\\
\bm v_{S^c}\in\mathbb{R}^{|S^c|},\; \bm u\ge0,\\
\bm b \in \{0,1\}^{s}
}}\quad &
\bm v^\top \hat{\Sigma} \bm v\;\mathrlap{\;\text{with }\bm{v}_S=\bm{v}_S^+ - \bm{v}_S^-}
\\
\text{subject to}\quad
& \sum_{j\in S} \bigl(v_j^+ + v_j^-\bigr) = 1, & \\
& v_j^+ \le M b_j        && (j\in S),\\
& v_j^- \le M(1-b_j)     && (j\in S),\\
& -u_j \le v_j \le u_j   && (j\in S^c),\\
& \sum_{j\in S^c} u_j \le 3. &
\end{alignedat}
\label{eq:miqp_comp}
\end{equation}
Alternatively, we may replace the Big-$M$ constraints $v_j^+ \le M b_j$ and $v_j^- \le M(1-b_j)$ in \eqref{eq:miqp_comp} with the SOS1 (special ordered set of type 1) constraint on $(v_j^+,v_j^-)$, which enforces that at most one of them is nonzero \citep{BealeTomlin1970,Williams2013}.

The mixed-integer problem can be handled by modern solvers such as Gurobi and is tractable even when $s$ is moderately large. In practice, however, the MIQP may require substantial computation time when $s$ is large. Therefore, we usually impose a time limit in our implementation so that the solver returns the best feasible solution within the time limit.

\section{Monte Carlo Simulation}
\subsection{Evaluation of compatibility constant via QP}
We investigate how the compatibility constant behaves and how tight the theoretical upper bound is under various sample sizes $n$ and dimensions $p$.  Because the compatibility condition strongly depends on the correlation structure of the design matrix, we generate data with different correlation structures.  The $p$-dimensional predictors are independently generated from $N_p(\bm{0}, \bm{\Sigma})$, where $\bm{\Sigma}$ has the compound-symmetry structure ${\Sigma} = (1-\rho) I_p + \rho \bm{1}_p\bm{1}_p^\top$.  The predictors are scaled to have mean zero and variances 1.  
The sample sizes are set to $n = 100, 200, 300, 400, 500, 750, 1000, 1500, 2000$, 
and the numbers of variables are set to $p = 20, 50, 100, 200, 500, 1000, 2000, 5000$.  The true coefficient vector $\bm{\beta}$ has $s=5$ nonzero elements, 
with indices $S \subseteq \{1, \dots, p\}$ chosen uniformly at random, and these nonzero elements are generated from $\mathrm{Unif}(1, 2)$. The response variables are generated from the regression model in \eqref{eq:regmodel} with $\bm \varepsilon \sim N(\bm 0, \sigma^2I_n),$ where $\sigma^2$ is determined such that the signal-to-noise ratio (SNR) is equal to 1; that is, $
  \sigma^2 := \bm{\beta}^\top \bm{\Sigma}\bm{\beta}$.  The regularization parameter is chosen as $\lambda = 2\sigma \sqrt{\frac{2}{n} \left(1+\log \left( \frac{p}{\delta} \right)\right)}$ to have the bound of \eqref{eq:error2}, where $\delta$ is set to 0.1. 

For each iteration of the Monte Carlo simulation, we generate a dataset and compute the compatibility constant $\phi$ by solving the QP with the OSQP solver using the \texttt{osqp} package in \texttt{R}. The lasso estimator is implemented with {\tt glmnet}. We obtain the following mean squared prediction error (MSPE) and its upper bound:
\begin{equation}
  \mathrm{MSPE}
  = \frac{1}{n}\|\bm{X}(\hat{\bm{\beta}} - \bm{\beta})\|_2^2,
  \qquad
  \mathrm{bound}
  = 9 \lambda^2 s / \phi^2.   \label{eq:evaluation} 
\end{equation}
Recall that the error bound in Proposition \ref{prop:lassoaccuracy} is proportional to $1/\phi^2$. Therefore, when the compatibility constant is close to zero, the theoretical upper bound becomes extremely large and practically uninformative. The computation of $\phi$ under different correlation structures and sample sizes helps us understand how the prediction error and its upper bound based on $\phi$ behave in finite samples.

\begin{figure}[t]
  \centering
  \includegraphics[width=\textwidth]{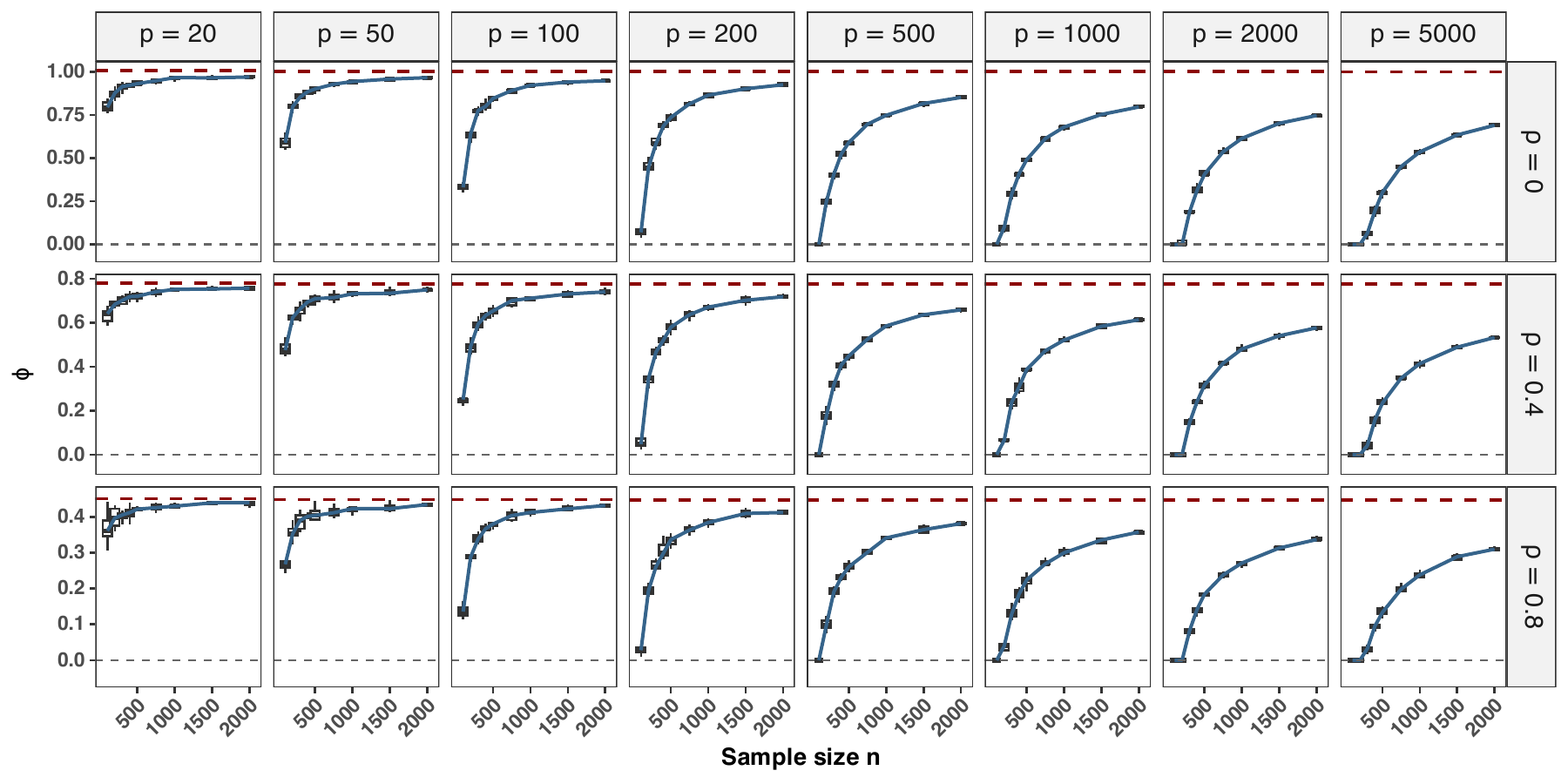}
  \caption{Compatibility constant $\phi$ for varying $(n, p, \rho)$. The red dashed line indicates the upper bound of compatibility constant for population covariance matrix, while the black dashed line indicates $\phi=0$.}
  \label{fig:phi}
\end{figure}
Figure \ref{fig:phi} presents the compatibility constant $\phi$ for different $n$, $p$, and $\rho$ under $R=10$ runs. The results show that the compatibility constant becomes larger as the number of observations increases. This behavior can be explained by the fact that the rank of the Gram matrix $\hat{\Sigma}=\sum_{i=1}^n\bm{x}_i\bm{x}_i^\top/n$ is less than or equal to $\min (n,p)$. When $n$ is small, the Gram matrix $\hat{\Sigma}$ has low rank and a large null space, allowing vectors satisfying the condition $\mathcal{C} := \{\|\bm{v}_{S^c}\|_1 \le 3\|\bm{v}_S\|_1\}$ to align with directions where the quadratic form $\bm v^\top \hat{\Sigma} \bm v$ becomes nearly zero. As $n$ increases, the rank of $\hat{\Sigma}$ increases and its null space shrinks, and as a result, it becomes harder to find such directions within $\mathcal{C}$.

We also observe that as $p$ increases, the compatibility constant decreases. This is probably because increasing $p$ enlarges the feasible set $\mathcal{C}$, which provides more degrees of freedom and makes it easier to find directions in which the quadratic form $\bm{v}^\top \hat{\Sigma}\bm{v}$ is small. In particular, when $p$ is large and $n$ is small, the compatibility constant sometimes becomes $\phi=0$, which implies the compatibility condition does not hold. Therefore, for high-dimensional data, several hundred observations would be necessary to get a reasonable compatibility constant. The compatibility constant also highly depends on the value of $\rho$; as the value of $\rho$ increases, the value of $\phi$ decreases. This can be explained by noting that the compatibility constant behaves similarly to the square root of the smallest eigenvalue of the covariance matrix, $\sqrt{1-\rho}$, which decreases as $\rho$ increases.

The red dashed line indicates the theoretical upper bound for the compatibility constant derived from the population covariance matrix, $\Sigma$; when $s$ is even, $\phi^{2}=1-\rho$, and when $s$ is odd, $\phi^{2}\le (1-\rho)\left(1+\frac{1}{s(p-s)}\right)$. For the derivation of these upper bounds, please refer to Appendix \ref{app:compatibility-compound-symmetry}. We observe that when $p$ is small, the compatibility constant converges to the upper bound as $n$ becomes large. However, when $p$ is large, the speed of convergence to the bound is slow. In other words, even under moderate correlation structures, the sample compatibility constant can remain far from the population-based bound because of finite-sample limitations rather than the population correlation itself.

\begin{figure}[t]
  \centering
  \includegraphics[width=\textwidth]{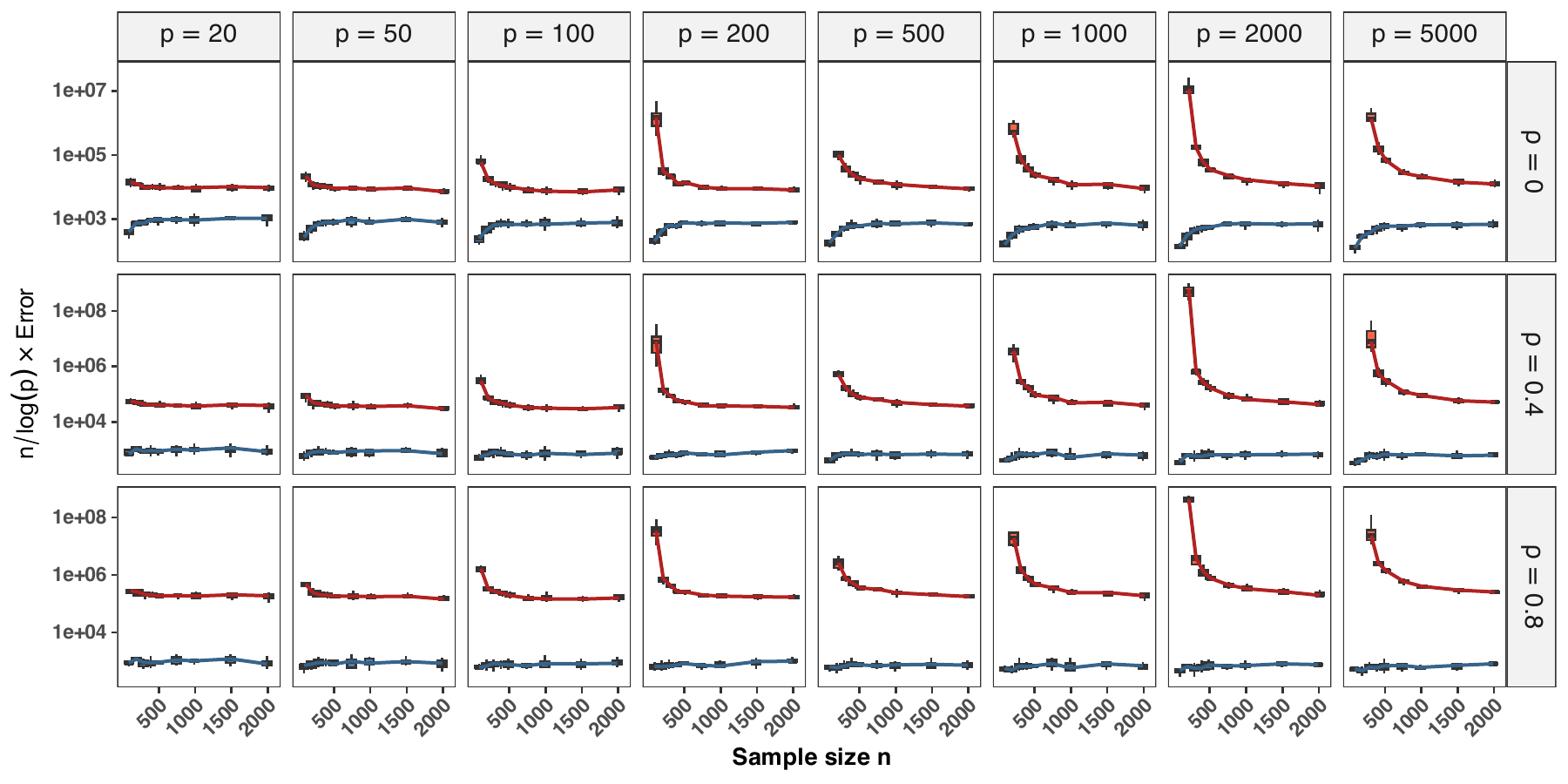}
  \caption{Scaled MSPE and upper bound, ${\rm MSPE}_{\rm scaled}$ and ${\rm bound}_{\rm scaled}$, for varying $(n, p, \rho)$.}
  \label{fig:collapse}
\end{figure}
Figure \ref{fig:collapse} shows the scaled MSPE and its upper bound as below
\begin{equation}
    {\rm MSPE}_{\rm scaled} = \frac{n}{\log p} {\rm MSPE}, \quad     
    {\rm bound}_{\rm scaled} = \frac{n}{\log p} {\rm bound},
    \label{eq:scaled}
\end{equation}
where the MSPE and its bound are defined in \eqref{eq:evaluation}.
 Here, the ``scaled" means multiplication of $n/\log p$ so that the right-hand side of inequality \eqref{eq:error2}, ${\rm MSPE} \lesssim C\log p /n$, has a quantity of order one. 

The result shows that the scaled MSPE is relatively stable regardless of the value of $n$ and $p$, except for small $n$.  On the other hand, the upper bound is unstable when $n$ is small and $p$ is large.  This is because the compatibility constant can be small for small $n$ as discussed in Figure \ref{fig:phi}.  In particular, with large $p$ small $n$, the upper bound cannot be obtained because of $\phi=0$.  Therefore, the upper bound made by the small $n$ can be much larger than that with large $n$.  

\begin{figure}[t]
  \centering
  \includegraphics[width=\textwidth]{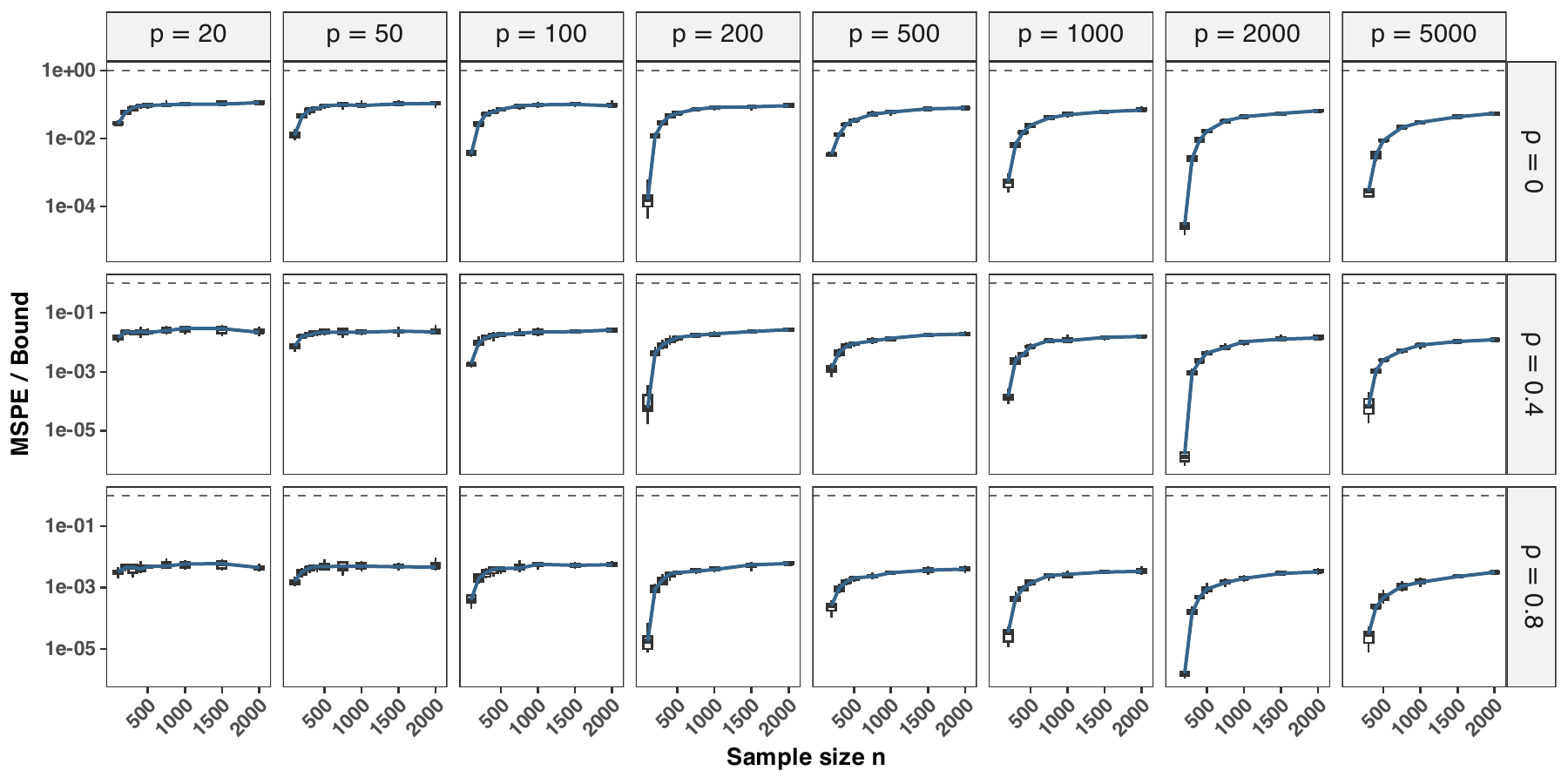}
  \caption{Ratio between the MSPE and the upper bound, $\text{MSPE}/\text{Bound}$, for varying $(n, p, \rho)$.}
  \label{fig:ratio}
\end{figure}
Figure \ref{fig:ratio} shows the ratio between the MSPE and its upper bound.  The ratio becomes extremely small for high-dimensional settings and small sample sizes.  In particular, the ratio can be $10^{-6}$ to $10^{-5}$, which implies that the upper bound does not provide a meaningful evaluation of the MSPE.  Therefore, the compatibility constant cannot be useful unless $n$ is relatively large.  For large sample sizes, the ratio is $10^{-2}$ to $10^{-1}$, which is reasonable to evaluate the MSPE but still has a gap between MSPE and bound.  As a result, when the compatibility constant is sufficiently large, the MSPE would be small so that the lasso might perform well.  Even if the compatibility constant is small, there exist situations where the MSPE is sufficiently small compared with its bound.

\subsection{Evaluation of the prediction error bound at each step in the proof of the oracle inequality}
In the previous simulation study, we observe that the gap between the MSPE and its bound is generally large, especially when $p$ is large. The gap comes from many inequalities obtained from each step of the proof in Proposition \ref{prop:lassoaccuracy}. The inequality becomes looser as the steps of the proof increase. Here, we evaluate the gap for each step of the proof, and investigate the key inequality that loosens the gap significantly.

\begin{table}[!t]
\centering
\caption{Bound of MSPE for each step of the proof. Here, $\bm \Delta := \hat{\bm\beta} - \bm \beta$.}
\label{tab:Proof step}
\begin{tabular}{c l l}
\hline
Step & Inequality & Bound \\
\hline
B
&
Basic inequality
&
$\displaystyle
\mathrm{MSPE}
\le
\frac{2}{n}\bm \varepsilon^\top X\bm \Delta
+
2\lambda\bigl(\|\bm \beta\|_1-\|\hat{\bm \beta}\|_1\bigr)
$
\\[1.2ex]

Du
&
Dual norm
&
$\displaystyle
\mathrm{MSPE}
\le
2\left\|\frac{X^\top\bm \varepsilon}{n}\right\|_\infty \|\bm \Delta\|_1
+
2\lambda\bigl(\|\bm \beta\|_1-\|\hat{\bm \beta}\|_1\bigr)
$
\\[1.2ex]

M
&
Max bounded by $\lambda/2$
&
$\displaystyle
\mathrm{MSPE}
\le
\lambda\|\bm \Delta\|_1
+
2\lambda\bigl(\|\bm \beta\|_1-\|\hat{\bm \beta}\|_1\bigr)
$
\\[1.2ex]

T
&
Triangle
&
$\displaystyle
\mathrm{MSPE}
\le
3\lambda\|\bm \Delta_S\|_1
-
\lambda\|\bm \Delta_{S^c}\|_1
$
\\[1.2ex]

Dr
&
Drop negative term
&
$\displaystyle
\mathrm{MSPE}
\le
3\lambda\|\bm \Delta_S\|_1
$
\\[1.2ex]

C
&
Compatibility
&
$\displaystyle
\mathrm{MSPE}
\le
3\lambda
\sqrt{\frac{s}{\phi^2}}
\sqrt{\mathrm{MSPE}}
$
\\[1.2ex]

F
&
Final
&
$\displaystyle
\mathrm{MSPE}
\le
\frac{9\lambda^2 s}{\phi^2}
$
\\
\hline
\end{tabular}
\end{table}

The proof of the oracle inequality in \eqref{eq:error} mainly consists of 7 steps. For detail, please refer to Appendix \ref{app:MSPEderivation}. Table \ref{tab:Proof step} shows the bounds made by each step of the inequality. First, we employ a basic inequality (referred to as ``B") constructed by the minimization of the empirical loss.  Next, the inequality $|\bm a^T\bm b| \leq \|\bm a\|_\infty \|\bm b\|_1$ for given two vectors, $\bm a$ and $\bm b$, is used to evaluate the inner product of error variables and the prediction values (referred to as ``Du" because it involves the dual inequality). The bound, $\left\|X^\top\bm \varepsilon\right\|_\infty$, made by the dual inequality is bounded by $n\lambda/2$ with probability at least $1-\delta$ (referred to as ``M", indicating Maximum bound).  We then employ the triangle inequality (referred to as ``T") and drop a negative term (referred to as ``Dr"). The compatibility condition (referred to as ``C") is used to evaluate the value of $\| \bm\beta_S - \hat{\bm \beta}_S \|_1$.  Finally, the oracle inequality is derived by squaring both sides of the inequality derived from the compatibility condition (referred to as ``F").

\begin{figure}[!t]
\centering
    \includegraphics[width=\textwidth]{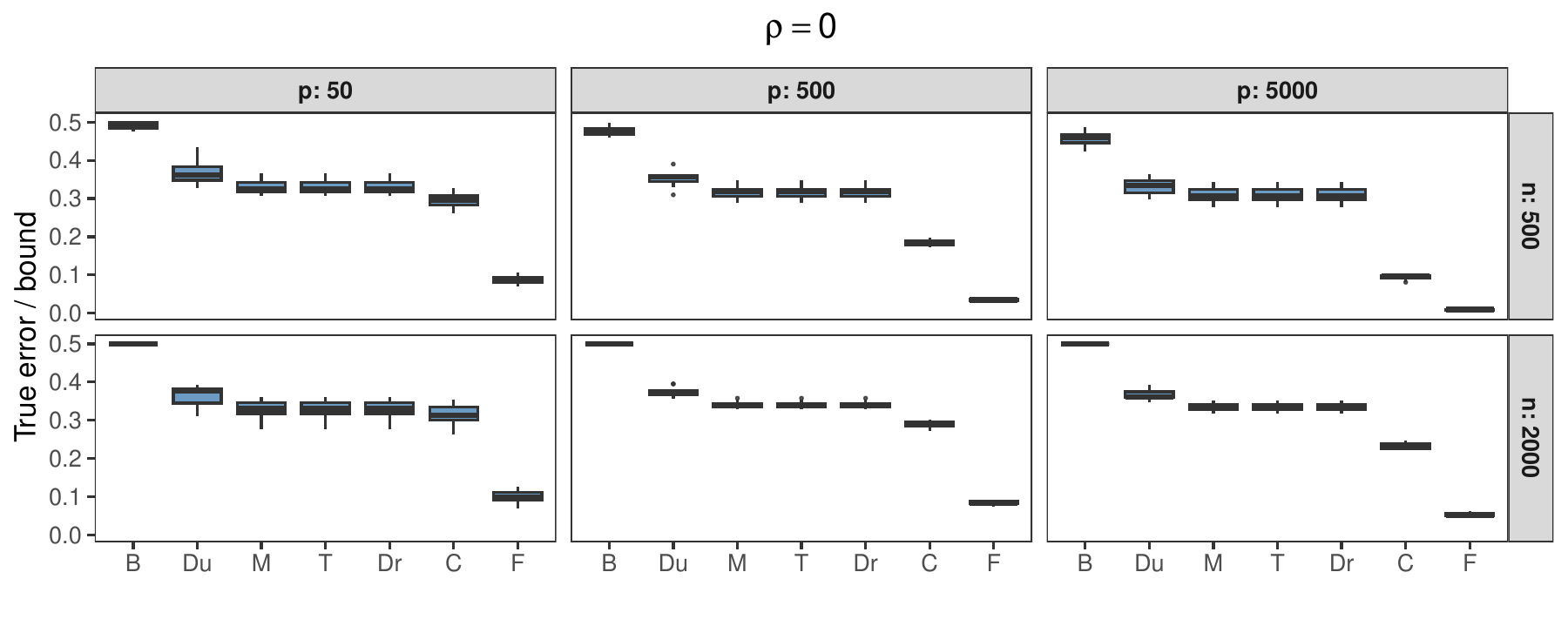}
\\
    \includegraphics[width=\textwidth]{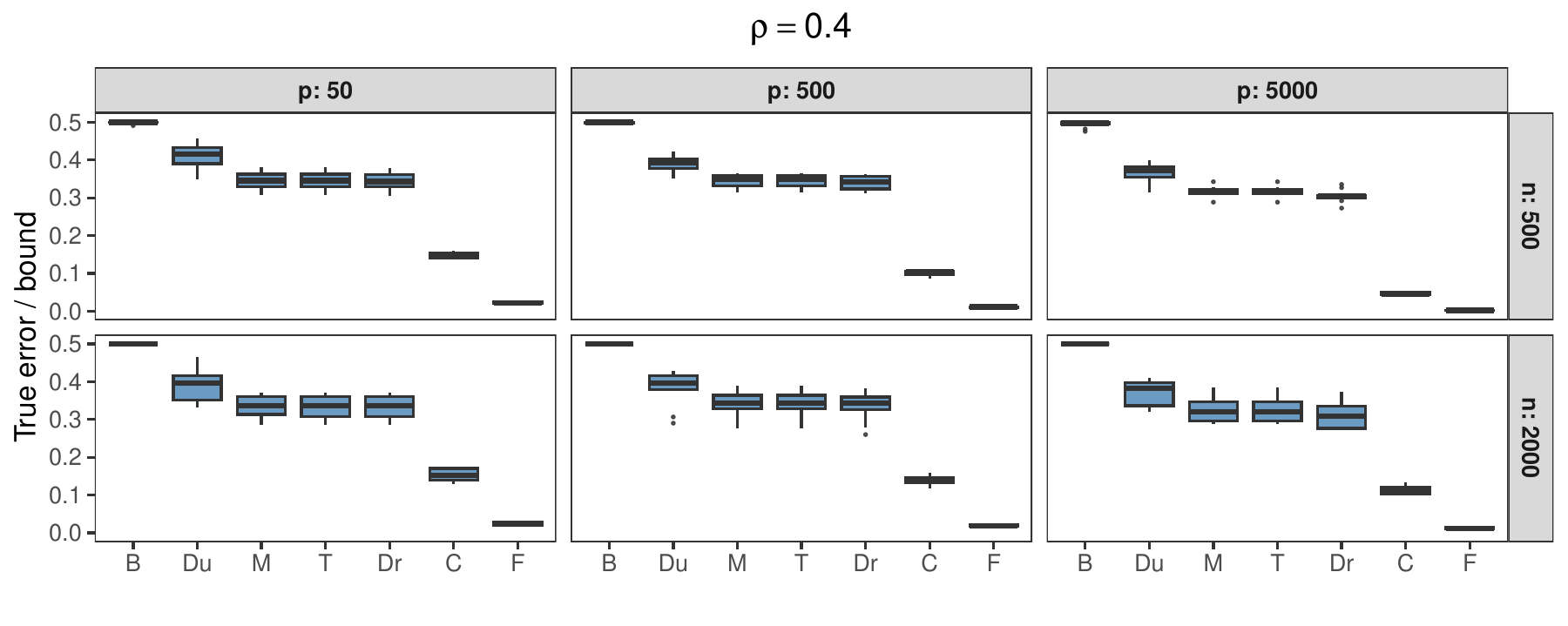}
\\
    \includegraphics[width=\textwidth]{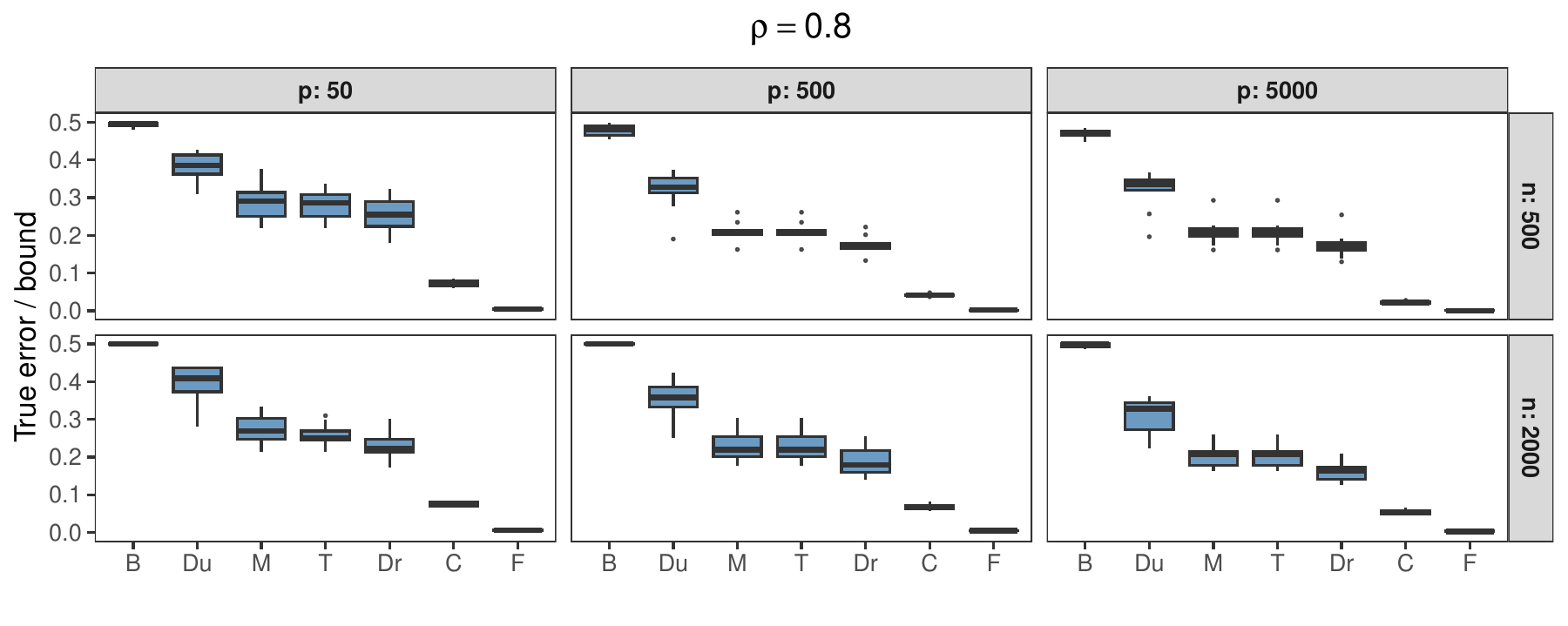}
\caption{Ratio between the actual MSPE and its bound for each step of the proof of inequalities. The seven steps are denoted by B (basic inequality), Du (dual norm), M (max inequality), T (triangle inequality), Dr (dropping the negative term), C (compatibility inequality), and F (final bound).}
\label{fig:audit_boxplots}
\end{figure}

Figure~\ref{fig:audit_boxplots} shows the ratio between the MSPE and its bound for each step of the inequalities. For clarity, we consider sample sizes $n = 500, 2000$, numbers of variables $p = 50, 500, 5000$, and correlation parameters $\rho = 0$, $0.4$, and $0.8$.  When evaluating the ratio, we first check the empirical frequency of the event, $\left\|X^\top \bm{\varepsilon} / n\right\|_\infty \le \lambda/2.$  Theoretically, this event occurs with probability at least $1-\delta = 0.9$, and we observe that it holds in all trials for almost all settings, except for $(n,p)=(2000,50)$ and $(2000,5000)$ with $\rho=0.4$, where the empirical frequency of this event is 0.9 over $R=10$ runs. The plots in Figure \ref{fig:audit_boxplots} are shown only for trials in which this event holds.

We obtain the following observations from Figure \ref{fig:audit_boxplots}:
\begin{itemize}
    \item The bounds of T (triangle) and Dr (drop the negative term) do not decrease much for all settings. The bound of M (max bounded by $\lambda/2$) theoretically decreases with probability at least $1-\delta$, and it actually decreases in many cases but does not decrease much, except for the large $\rho$, large $p$ case.  
    \item B, Du, C, and F can result in a substantial decrease in the ratio. Thus, the looseness of the final bound is attributed to some of B, Du, C, and F.
    \item For any cases, the ratio at B is often around one half. Indeed, the ratio is exactly the half when OLS (ordinary least squares), $\hat{\bm\beta}^{\rm OLS}$, is used.  As \(\bm y-X\hat{\bm\beta}^{\rm OLS}\) and \(X(\hat{\bm\beta}^{\rm OLS}-\bm\beta)\) are orthogonal, we have $\|\bm y-X\bm\beta\|^2=\|\bm y-X\hat{\bm\beta}^{\rm OLS}\|^2+\|X(\hat{\bm\beta}^{\rm OLS}-\bm\beta)\|^2$. Consequently, ${\rm MSPE} = \bm \varepsilon ^\top X(\hat{\bm\beta}^{\rm OLS}-\bm\beta)/n$. The half ratio suggests that the lasso estimator behaves similarly to OLS, in the sense that \(\bm y-X\hat{\bm\beta}\) and \(X(\hat{\bm\beta}-\bm\beta)\) are nearly orthogonal.
    \item When $p$ and $\rho$ are small and $n$ is large, the bound made by the compatibility condition (C) is reasonably tight. However, outside such settings, the drop with C becomes much worse.  
    \item In all cases, the ratio drops substantially at F in the end. The final step removes the square root by squaring both sides, which substantially enlarges the bound. 
\end{itemize}

\subsection{Computation of upper bound via MIQP}
The QP with all sign combinations described above can be applied only when $s$ is sufficiently small, such as $s=5$. When $s$ is large, such as $s=50$, the QP approach becomes infeasible even with parallel computation. In this case, mixed-integer quadratic programming (MIQP) provides a more feasible alternative.

First, we consider a small problem for which the QP can be computed within a realistic time, and compare the results with those from MIQP. For MIQP, we employ warm starts and early stopping with a time limit to evaluate the upper bound of the compatibility constant, and investigate how accurate this upper bound is.

\begin{figure}[!t]
\centering
\includegraphics[width=\linewidth]{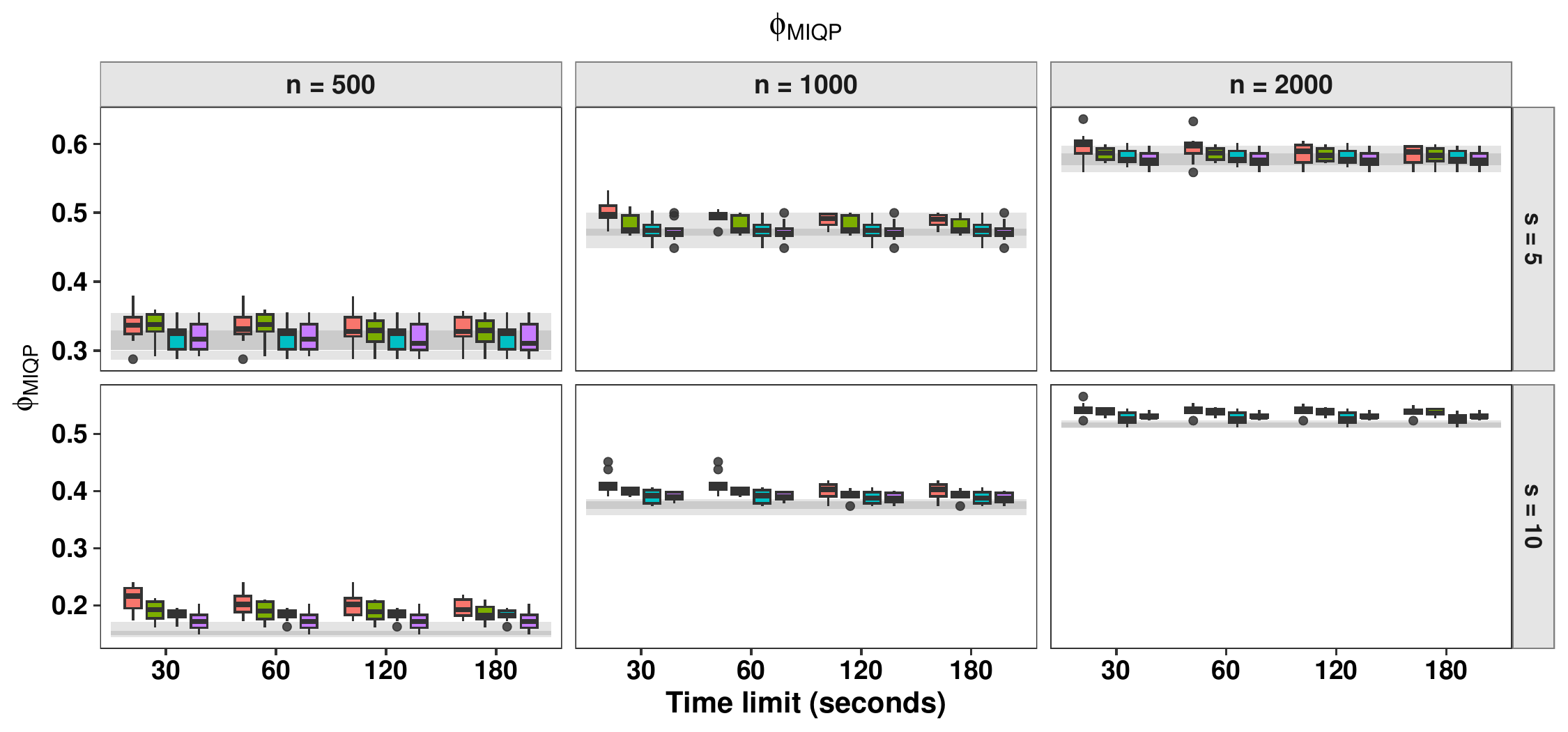}
\\
\includegraphics[width=\linewidth]{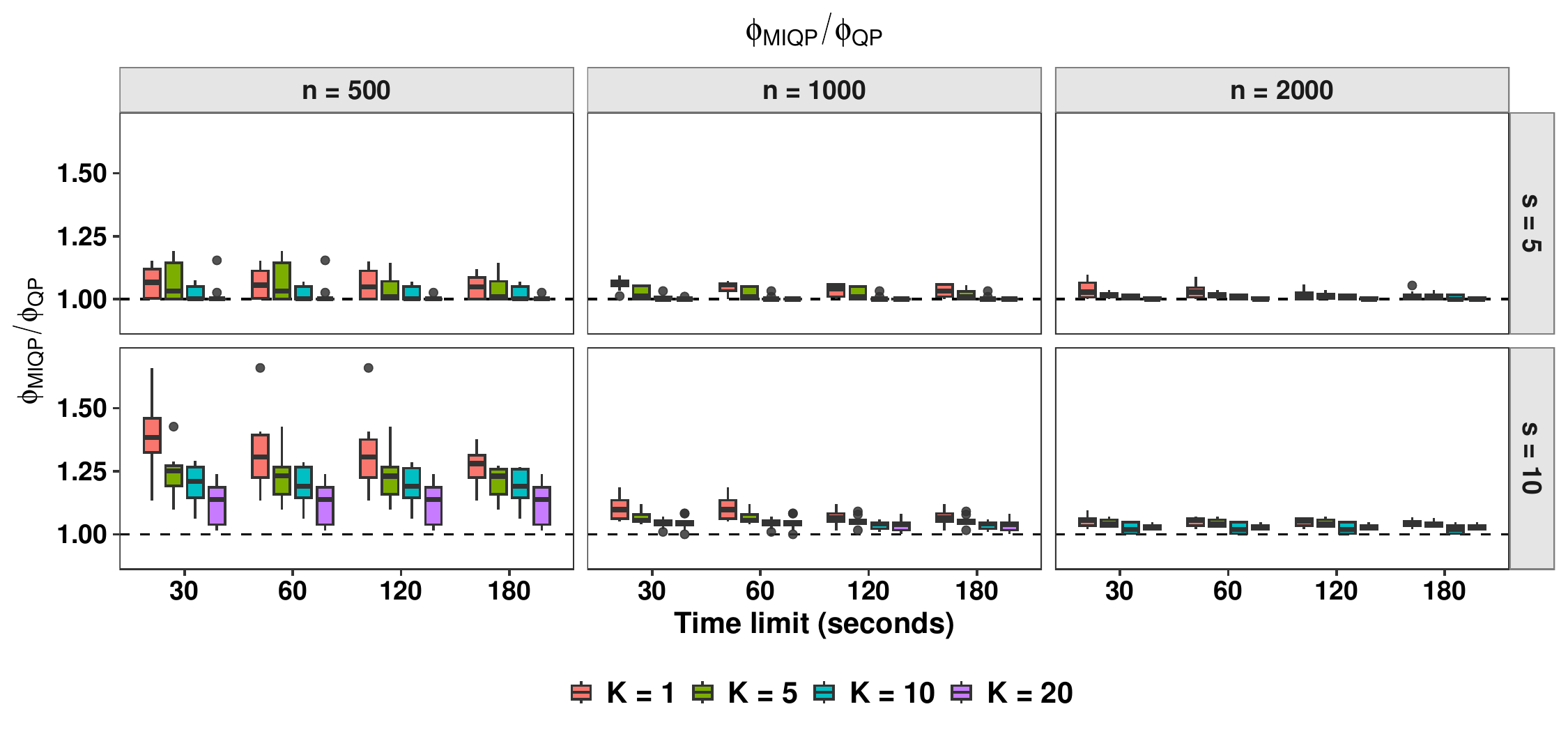}
\caption{
Compatibility constant obtained by MIQP and QP (upper panel) and their ratio, $\phi_{\mathrm{MIQP}} / \phi_{\mathrm{QP}}$ (lower panel). In the upper panel, the gray shaded regions summarize the distribution of the compatibility constant obtained by QP: the dark gray band represents between the first and third quartiles, while the light gray bands show the overall ranges from the minimum to the maximum. The boxplots show the MIQP-based estimates for different warm-start sizes $K$ and time limits.
}
\label{fig:timelimit_ratio}
\end{figure}

Figure \ref{fig:timelimit_ratio} shows the compatibility constant obtained by MIQP and QP (upper panel) and their ratio, $\phi_{\mathrm{MIQP}} / \phi_{\mathrm{QP}}$ (lower panel). In the upper panel, the gray shaded regions summarize the distribution of the compatibility constant obtained by QP: the dark gray band represents between the first and third quartiles, while the light gray bands show the overall ranges from the minimum to the maximum. The boxplots show the MIQP-based estimates for different warm-start sizes $K$ and time limits. The warm start value is obtained by solving the QP with a fixed sign vector $\bm{z}$. A set of $K$ ($K=1, 5, 10, 20$) random sign vectors $\bm{z}$ is generated from a Bernoulli distribution. For each vector, we solve the corresponding QP and select the one that yields the smallest objective value. The time limits for MIQP are set to $T = 30, 60, 120,$ and $180$ seconds. The other parameters are set to $p=2000$ (i.e., relatively high-dimensional case), $\rho=0.4$, $n = 500$, $1000$, $2000$, and $s=5$ and $10$.

For all cases, the compatibility constant obtained by MIQP is always larger than that obtained by QP because MIQP uses a time limit and cannot always attain the optimal value. The difference between MIQP and QP is relatively small, especially for large $n$. When $n$ is small, MIQP can yield a relatively larger $\phi$ than QP, and thus the difference is not negligible. Nevertheless, the ratio between the two compatibility constants, $\phi_{\mathrm{MIQP}} / \phi_{\mathrm{QP}}$, never exceeds 2, implying that the order of the error bound may not change due to the difference between MIQP and QP.

The ratio does not change substantially across different time limits for a fixed $K$, indicating that the warm start provides a reasonable initial value. We also observe that the effect of $K$ is much larger than that of the time limit; as $K$ increases, the ratio decreases substantially. These findings suggest that most of the improvement is achieved by the warm start based on $K$ random signs, and MIQP has only a limited search space left to explore. Therefore, increasing the time limit gives almost no additional benefit, while increasing $K$ significantly enhances the quality of the warm start. It would be preferable to choose a large $K$, while the time limit does not necessarily have to be large; 60 seconds appears to be sufficient when $K$ is large.

\begin{figure}[t]
\centering
\includegraphics[width=\textwidth]{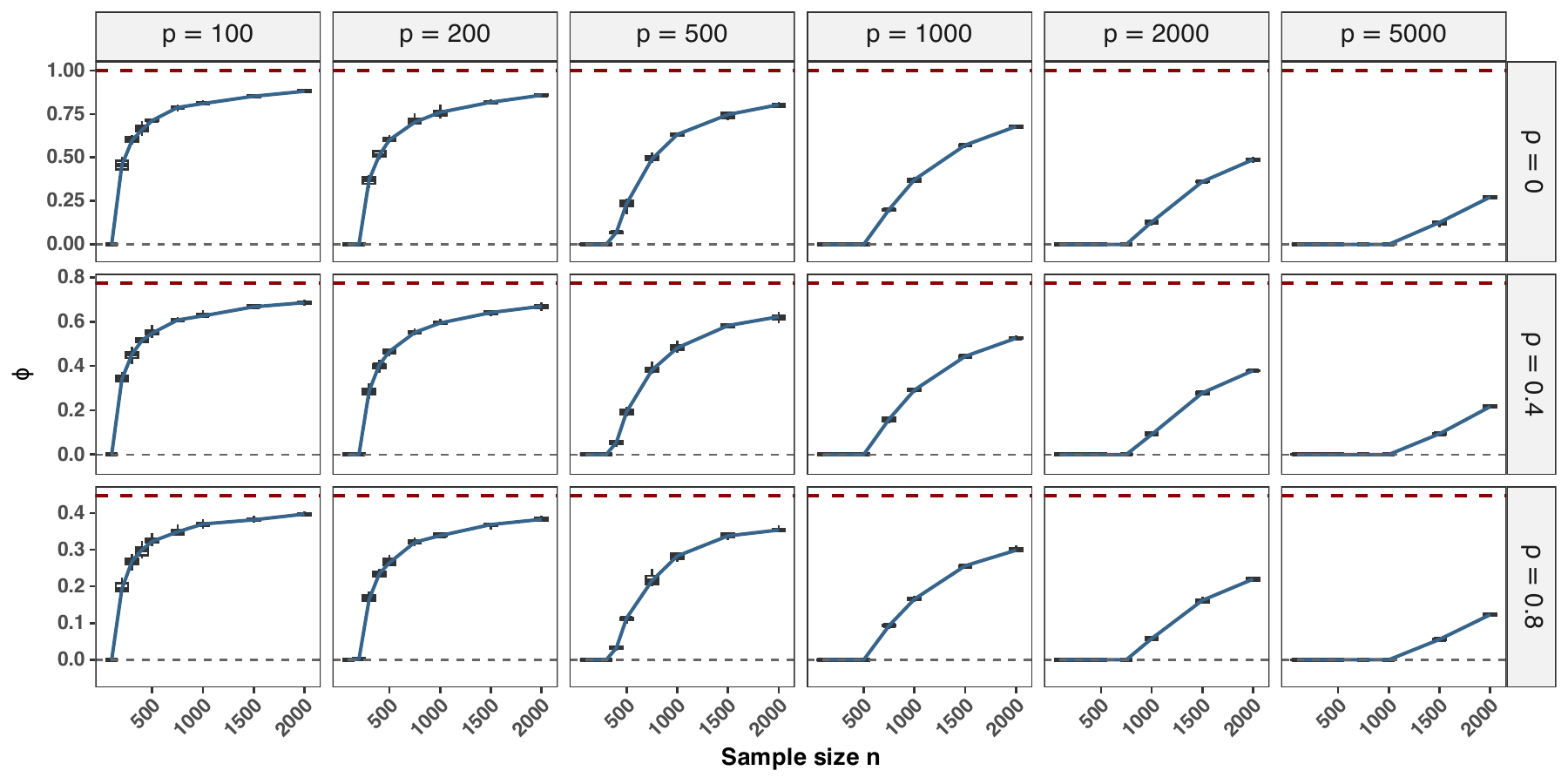}
\caption{Compatibility constant $\phi$ with $s=50$ for varying $(n, p, \rho)$. The red dashed line indicates the compatibility constant for the population covariance matrix, $\sqrt{1-\rho}$, while the black dashed line indicates $\phi=0$.}
\label{fig:phi_miqp}
\end{figure}

Next, we consider a larger problem with $s=50$, where the QP with all sign combinations cannot be performed in practice due to the high computational cost of $2^{50-1}$ patterns. We compute an upper bound of $\phi$ using MIQP described above. Based on Figure \ref{fig:timelimit_ratio}, we set $K=20$ and a time limit of $T=60$ seconds. The other settings are the same as in Figure \ref{fig:phi}.
 The results are shown in Figure \ref{fig:phi_miqp}. The red dashed line indicates the compatibility constant for the population covariance matrix, $\sqrt{1-\rho}$, while the black dashed line indicates $\phi=0$. 

The results show a similar tendency to Figure \ref{fig:phi}, but yield much lower $\phi$ for small sample sizes. In particular, when $p=5000$ and $n=1000$ (high-dimensional and relatively large sample sizes), the upper bound of compatibility constant is $0$, indicating that the compatibility constant is also $0$. Therefore, even when $n$ is relatively large, the compatibility constant can be zero or close to zero. This result shows a clear finite-sample gap compared with asymptotic intuition. More surprisingly, even in the uncorrelated case ($\rho = 0$), the compatibility condition does not hold simply due to the relationship among $(n,p,s)$. Therefore, the finite-sample behavior is strongly dominated by these three factors rather than true correlation among predictors. The MIQP with time limit cannot always be exact, but is helpful here because it can detect that $\phi$ becomes zero or close to zero in challenging settings. 

\section{Real data analysis}
We investigate how the compatibility constant $\phi$ and the error bound behave as the sample size increases through the analysis of real data. We use S\&P 500 constituents to investigate the relationship between the daily return of Apple and that of the remaining index members.   Since the composition of the S\&P 500 index changes over time, not all the 500 stocks are continuously observed over the entire period.  We download data from Yahoo Finance for the period from November 2017 to October 2025, and retain only the stocks for which daily records are available throughout this period.  This screening results in 476 stocks, one of which is Apple.

In the regression analysis, the daily return of Apple is treated as the response variable, while the returns of the remaining 475 stocks are used as predictors.  We consider the arithmetic daily returns
\[
  R_{t,j} = \frac{P_{t,j} - P_{t-1,j}}{P_{t-1,j}},
\]
where $P_{t,j}$ denotes the closing price of stock $j$ on day $t$. The data set consists of $n = 2009$ observations and $p = 475$ predictors. In what follows, the predictors are standardized and the response vector $\bm y$ is centered.

\begin{figure}[t]
  \centering
  \includegraphics[width=\textwidth]{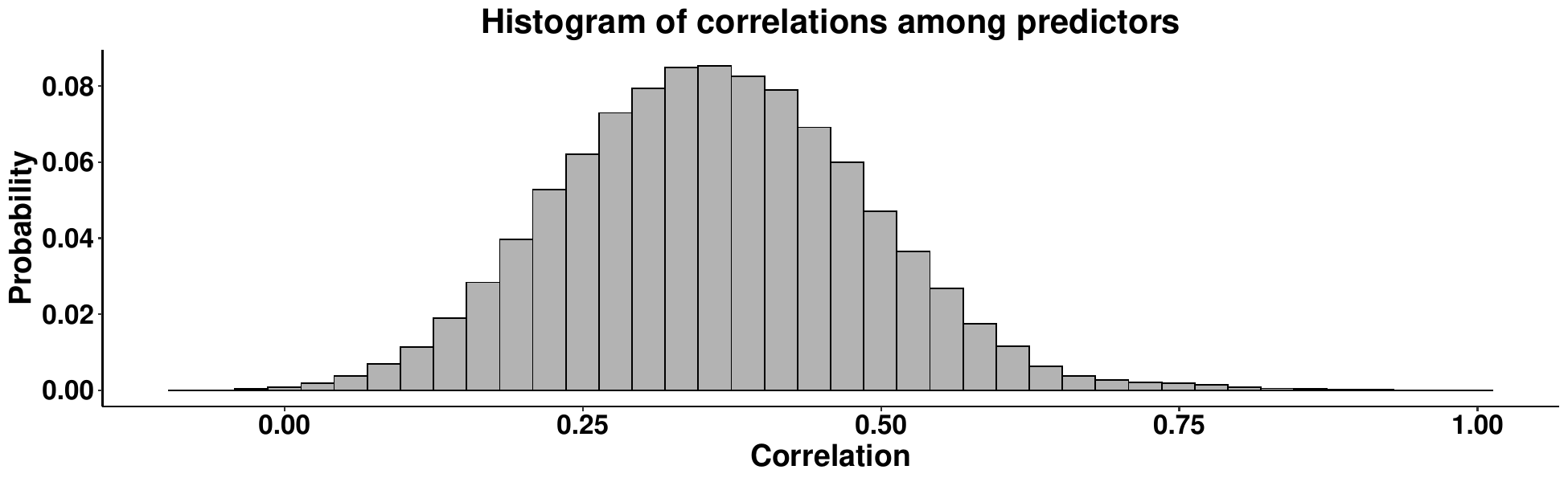}
  \caption{%
    Histogram of correlations among predictors.  }
  \label{fig:histogram_cor}
\end{figure}
Figure \ref{fig:histogram_cor} shows the histogram of correlations among predictors. Most pairs of variables are moderately and positively correlated, while a few exhibit relatively high correlations.
This correlation structure may influence the value of the compatibility constant, although its effect appears to be limited.

To compute the compatibility constant, we need the support of the true coefficients, which cannot be provided in practice.  Therefore, we provide a heuristic method to estimate the support of the true coefficients. First, we divide the dataset into training and test datasets. The number of observations for the training and test data is $n_{\rm train}=1009$ and $n_{\rm test}=1000$, respectively.  Then, we fit the lasso and estimate the set $S$ as follows:
\begin{enumerate}
\item We first select the regularization parameter $\lambda$ by the 10-fold cross validation.  The cross-validation is known to be inconsistent in model selection and tends to select overly complex models \citep{shao1993linear}.  Indeed, the number of nonzero coefficients selected by the cross-validation is $77$, which is relatively large.                
  \item 
  We obtain an estimator of $\sigma^2$ defined by $\hat\sigma^2  = \frac{1}{n_{\mathrm{train}} - s_{\mathrm{CV}} - 1} \bigl\|\bm y_{\mathrm{train}} - X_{\mathrm{train}}\hat{\bm\beta}^{(\mathrm{CV})}\bigr\|_2^2$. Numerically,  $\hat\sigma^2 = 1.268\times 10^{-4}.$
  \item Based on the above $\hat\sigma^2$, we estimate the penalty level based on the Proposition \ref{prop:lassoaccuracy}
        by
        \begin{equation}\label{eq:lambda-train}
          \lambda_{\mathrm{train}}
          = 2\hat\sigma
            \sqrt{\frac{2}{n_{\mathrm{train}}}
              \Bigl(1 + \log\frac{p}{\delta}\Bigr)}
        \end{equation}
        with $\delta = 0.1$.  We get $\lambda_{\mathrm{train}} = 3.085\times 10^{-3}.$ 
\item We fit a lasso  on
        $(X_{\mathrm{train}}, \bm y_{\mathrm{train}})$ again but with the above $\lambda_{\mathrm{train}}$.  With the estimated regression coefficients, say
        $\hat{\bm\beta}^{(\mathrm{train})}$, we get a set of nonzero coefficients,
          $\hat{S}
          = \operatorname{supp}(\hat{\bm\beta}^{(\mathrm{train})})
        $. 
\end{enumerate}
 In our experiment, we obtain the following 
$|\hat{S}| = 11$ companies:  Microsoft Corp., Amazon.com Inc., Meta Platforms Inc.,
Berkshire Hathaway Inc.\ (Class~B), Mastercard Inc.,
Cisco Systems Inc., Amphenol Corp., IDEXX Laboratories Inc.,
Fastenal Co., CDW Corp., and Skyworks Solutions Inc.

The estimate of the active set, $\hat{S}$, can be used to compute the compatibility constant with the test dataset.  To investigate the behavior of the compatibility constant with various sample sizes, we use the first $n = 100k$ ($k=1,2,\dots,10$) observations of the test block. Denote the resulting subsample by $(X_n, y_n)$. For each $n$, we compute the empirical compatibility constant of $X_n$ with
respect to the fixed set $\hat{S}$, say $\phi_n = \phi\bigl(\hat{S}; X_n\bigr)$, using QP.   

We also estimate the MSPE and its bound computed with the $\phi_n$ to obtain the results similar to the Figures \ref{fig:phi}--\ref{fig:ratio} in the simulation study.  For a given $n$, we define the regularization parameter as
$  \lambda_n
  = 2\hat\sigma
    \sqrt{\frac{2}{n}
      \Bigl(1 + \log\frac{p}{\delta}\Bigr)},
  $ with $\delta = 0.1$, 
which is the analog of~\eqref{eq:lambda-train} with $n$ in place of
$n_{\mathrm{train}}$.  We then compute the lasso estimate, $\hat{\bm\beta}_n$, with the above regularization parameter.  The MSPE in \eqref{eq:evaluation} is then estimated as
\[
  \widehat{\mathrm{MSPE}}
  =
  \frac{1}{n}\bigl\|X_n(\hat{\bm\beta}_n - \hat{\bm{\beta}}^{\rm train})\bigr\|_2^2, \quad   \mathrm{Bound}_n
  := \frac{9\,|\hat{S}|\,\lambda_n^2}{\phi_n^2}.
\]
Here, the true $\bm \beta$ and $S$ are replaced by their estimates with training data, $\hat{\bm{\beta}}^{\rm train}$ and $\hat{S}$, respectively.  

\begin{figure}[t]
  \centering
  \includegraphics[width=\textwidth]{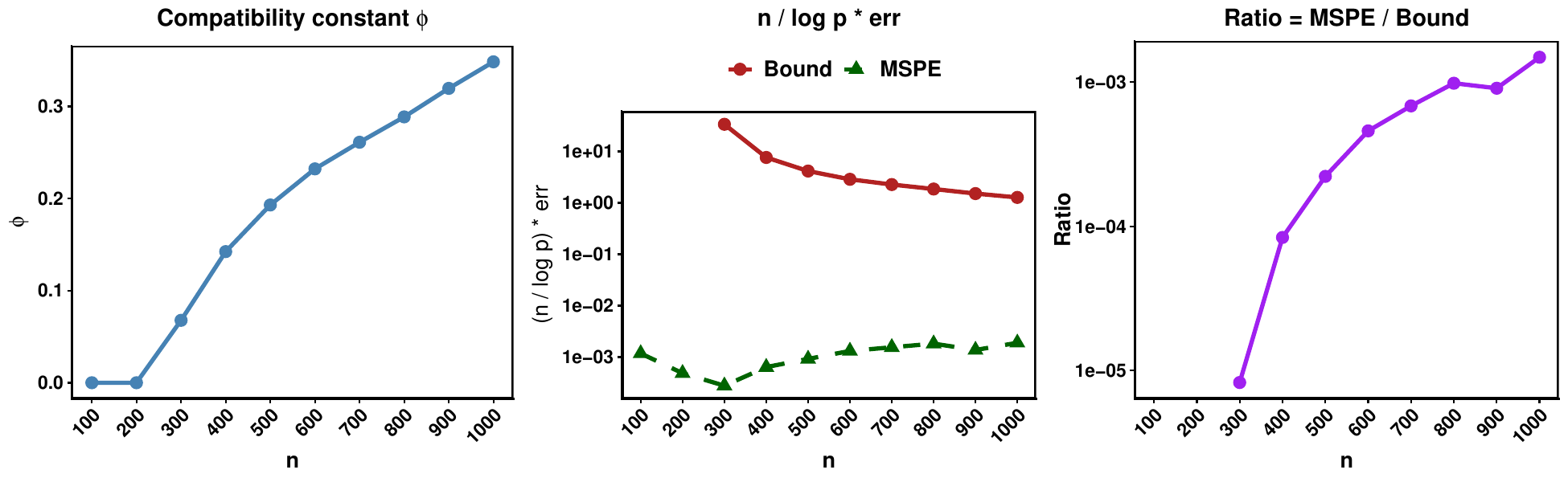}
  \caption{The estimate of compatibility constant $\phi_n$ (left panel), the MSPE and its error bound scaled by $n/\log p$ as described in \eqref{eq:scaled} in the simulation study (middle panel), and their ratio, $  \mathrm{Ratio}_n =\widehat{\mathrm{MSPE}}/ {\mathrm{Bound}_n}$ (right panel).}
  \label{fig:realdata-sp500}
\end{figure}

Figure \ref{fig:realdata-sp500} shows the estimate of compatibility constant $\phi_n$, the MSPE and its error bound scaled by $n/\log p$ as described in \eqref{eq:scaled} in the simulation study, and their ratio, $  \mathrm{Ratio}_n =\widehat{\mathrm{MSPE}}/ {\mathrm{Bound}_n}$.
The estimated compatibility constant is a monotone non-decreasing function of $n$, which implies a large sample size is required to achieve a good error bound.  For $n=100$ and $n=200$, $\phi_n=0$ and then the compatibility condition may not hold.  Meanwhile, for large $n$, the compatibility constant can be greater than $0.3$, but it is still not very large.  Considering that there exist a few pairs of large correlations in predictors, as shown in Figure \ref{fig:histogram_cor}, the moderate value of $\phi_n$ seems reasonable. 

The scaled $\widehat{\mathrm{MSPE}}$ is almost constant as a function of $n$, while the bound is a monotone decreasing function, which aligns with the simulation results.  This monotonicity is likely due to the fact that the compatibility constant is non-decreasing in $n$. The error bounds for $n=100$ and $200$ cannot be obtained because the compatibility condition does not hold.  The ratio is a monotone increasing function, which also aligns with the results from simulation study. The ratio is not small but not too large compared to the results from the simulation study (see Figure \ref{fig:ratio}).  As a result, the overall results almost align with the simulation study and may provide relatively good upper bound when a sufficient number of observations is available.

\section{Concluding remarks}
We have provided an approach to compute the compatibility constant when the support of the true regression coefficients is known.  The computation of the compatibility constant reduces to a QP if the signs of the true nonzero regression coefficients are given.  Thus, we can compute the compatibility constant by solving QP for all combinations of signs.  When the number of nonzero coefficients $s$ is large, we may alternatively use the MIQP with Big-$M$ or SOS1 under a time limit to obtain an upper bound of the compatibility constant.  A more efficient computation for large $s$ would be crucial, but we will leave this as a future research topic.

The compatibility condition is one of the most important conditions in the lasso literature for achieving good prediction accuracy in high-dimensional settings. Although the theoretical upper bound is useful, it can be conservative in finite samples.  Therefore, if we are interested in the finite-sample behavior of the lasso, both the condition and its implications should be interpreted carefully.  More generally, researchers are interested in developing strong theoretical properties of statistical methods under conditions that are as mild as possible. In many cases, however, these conditions cannot be evaluated with a given dataset due to the fact that the true parameter is unknown and the associated optimization problem is often nonconvex. Although exact verification is impossible in practice, it is still meaningful to consider numerical approaches to assess whether the regularity conditions are likely to hold for a given dataset.  The evaluation of regularity conditions would be useful in many practical situations, as it helps determine whether the method is expected to perform well. In the future, we would like to develop numerical approaches to evaluate the regularity conditions for various statistical methods.

\section*{Acknowledgments}
This work was supported by JSPS KAKENHI Grants JP23K11007, JP23H00466, JP23K22410, and JP23K01333.

\bibliographystyle{apalike}
\bibliography{papers}

\appendix

\section{Compatibility constant under compound symmetry}
\label{app:compatibility-compound-symmetry}

We study the compatibility constant for a covariance matrix that has a compound symmetry, expressed as
\[
 \Sigma_{\rho}
  := (1-\rho) I_p + \rho \mathbf{1}_p\mathbf{1}_p^\top,
  \qquad 0 \leq \rho < 1.
\]
Here, the correlation coefficient $\rho$ can be negative value, but for simplicity, we only consider the case where $\rho \geq 0$. Recall that the compatibility constant is defined as \eqref{eq:problem2}.  For any $\bm{v}\in\mathbb{R}^p$, we have
\begin{equation}
  \label{eq:quadratic-form-decomposition}
  \bm{v}^\top\Sigma_{\rho}\bm{v}
  =
  (1-\rho)\|\bm{v}\|_2^2
  +
  \rho\,(\mathbf{1}^\top\bm{v})^2
  =
  (1-\rho)\|\bm{v}\|_2^2
  +
  \rho\left(\sum_{j=1}^p v_j\right)^2.
\end{equation}
The compatibility constant is closely related to the smallest eigenvalue of the covariance matrix. The eigenstructure of $\Sigma_{\rho}$ is simple: there exists a unique maximum eigenvalue $\lambda_{\max} = 1 + (p-1)\rho$ with the eigenvector being $\mathbf{1}_p$, and for the other $(p-1)$-dimensional vectors that is orthogonal to $\bm{1}$, we have the same eigenvalues $\lambda_{\min} = 1-\rho$.  Thus, the quadratic form $\bm{v}^\top\Sigma_{\rho}\bm{v}$ is smallest
when $\bm{v}$ lies close to the subspace orthogonal to $\mathbf{1}_p$,
that is, when $\sum_{j=1}^p v_j \approx 0$. This is also reasonable from Eq.  \eqref{eq:quadratic-form-decomposition} as the second term penalizes the component along $\mathbf{1}_p$.

If we minimize $\bm{v}^\top\Sigma_{\rho}\bm{v}$ with $\|\bm{v}\|_2=1$,
it would simply mean choosing $\bm{v}$ as an eigenvector corresponding
to the smallest eigenvalue, $1-\rho$. However, the constraint on the compatibility constant uses $\|\bm{v}_S\|_1=1$
rather than $\|\bm{v}\|_2=1$.
Therefore, the decomposition of the quadratic form in \eqref{eq:quadratic-form-decomposition} implies that we want to minimize $\|\bm{v}\|_2^2$ under the $\ell_1$-constraint $\|\bm{v}_S\|_1=1$, while we also want $\sum_{j=1}^p v_j$ to be as close to $0$ as possible to avoid the direction corresponding to the largest eigenvalue.

We first derive a lower bound of $\phi$. Eq. \eqref{eq:quadratic-form-decomposition} is evaluated with 
\begin{equation}
s\bm{v}^\top\Sigma_{\rho}\bm{v}
    \geq s(1-\rho)\|\bm{v}\|_2^2
    \geq s(1-\rho)\|\bm{v}_S\|_2^2
    \geq (1-\rho)\|\bm{v}_S\|_1^2
    = (1-\rho).\label{eq:lower}    
\end{equation}
The last inequality is derived from Cauchy-Schwarz inequality with two vectors, $\bm{v}_S$ and ${\rm sign}(\bm{v}_S)$. Eq. \eqref{eq:lower} implies that the lower bound of $\phi^2$ is $1-\rho$. 

If the upper bound of $\phi^2$ is also $1-\rho$, the compatibility constant is $\phi^2=1-\rho$. Fortunately, it is true when $s$ is even.  When $s$ is odd, the derivation of exact value of $\phi^2$ would be more challenging. Therefore, we obtain an upper bound that is slightly greater than $1-\rho$. 

\begin{proposition}   
It is shown that
  \begin{align*}
\begin{alignedat}{2}
\phi^{2} &= 1-\rho 
    &\quad &(s:\ \text{even}),\\
\phi^{2} &\le (1-\rho)\left(1+\frac{1}{sr}\right)
    &\quad &(s:\ \text{odd}),
\end{alignedat}
\end{align*}
where $r := p-s$.
\end{proposition}
\begin{proof}
  Without loss of generality, we can define $S=\{1,2,\dots,s\}$ and $S^C=\{s+1,\dots,p\}$. Firstly, we assume that $s$ is even. Define $\bm{v}$ by
\begin{equation}
      v_j =
    \begin{cases}
      +\frac{1}{s}, &  j =1,\dots, \frac{s}{2},\\[2pt]
      -\frac{1}{s}, & j = \frac{s}{2}+1,\dots,s,\\[2pt]
      0, & j\in S^c.
    \end{cases}  
    \label{eq:beta_even}
\end{equation}
  The vector $\bm v$ satisfies the constraint in \eqref{eq:problem2} because $
    \|\bm v_S\|_1= 1$ and $\|\bm v_{S^c}\|_1 = 0 \le 3$. 
    Moreover, since $\sum_{j=1}^pv_j = 0$ and considering $\|\bm v\|_2^2 = \frac{1}{s}$, we obtain
  \[
    s\,\bm v^\top\Sigma_{\rho}\bm v = 1-\rho.
  \]
Therefore, $\phi^2=1-\rho$ when $s$ is even.

Next, we derive an explicit upper bound of $\phi$ when $s$ is odd. It is difficult to define $\bm{v}$ as in \eqref{eq:beta_even} because $s$ is odd, but we may define $\bm v$ that is similar to that in the even case in \eqref{eq:beta_even}:
\begin{equation}
      v_j =
    \begin{cases}
      +\frac{1}{s}, &  j =1,\dots, \frac{s-1}{2},\\[2pt]
      -\frac{1}{s}, & j = \frac{s-1}{2}+1,\dots,s,\\[2pt]
      \frac{1}{sr}, & j\in S^c.
    \end{cases}  
    \label{eq:beta_odd}
\end{equation}
  The vector $\bm v$ satisfies the constraint in \eqref{eq:problem2} because $
    \|\bm v_S\|_1= 1$ and $\|\bm v_{S^c}\|_1 = 1/s \le 3$. Furthermore, $\bm v_{S^c}$ is selected so that $\sum_{j=1}^pv_j=0$ is satisfied. Since $\|\bm v_{S^c}\|_2^2=1/(s^2r)$, we have
  \[
    s\,\bm v^\top\Sigma_{\rho}\bm v
    =
    s(1-\rho)\left(\frac{1}{s} + \frac{1}{s^2 r}\right)
    =
    (1-\rho)\left(1 + \frac{1}{s r}\right),
  \]
  which is the upper bound for $\phi^2$
\end{proof}

\section{Derivation of the MSPE bound in Proposition \ref{prop:lassoaccuracy}}\label{app:MSPEderivation}

Here, we derive the upper bound in Proposition \ref{prop:lassoaccuracy} step by step and clarify which inequality is used at each stage. Our goal is to derive
\begin{equation*}
\frac{1}{n}\|X\hat{\bm{\beta}}-X\bm{\beta}\|_2^2
\le
\frac{9\lambda^2 s}{\phi^2},
\label{eq:goal}
\end{equation*}
where $\phi$ denotes the compatibility constant.

\paragraph{Step 1: Basic inequality.}

Since $\hat{\bm{\beta}}$ minimizes the lasso objective function, comparing the objective values at $\hat{\bm{\beta}}$ and $\bm{\beta}$ yields
\begin{equation}
\frac{1}{2n}\|\bm{y}-X\hat{\bm{\beta}}\|_2^2 + \lambda \|\hat{\bm{\beta}}\|_1
\le
\frac{1}{2n}\|\bm{y}-X\bm{\beta}\|_2^2 + \lambda \|\bm{\beta}\|_1.
\label{eq:basic1}
\end{equation}
Eq. \eqref{eq:basic1} is referred to as the \emph{basic inequality}. Let $\bm{\Delta}$ be
\begin{equation}
\bm{\Delta} = \hat{\bm{\beta}} - \bm{\beta}.
\label{eq:Delta}
\end{equation}
Rearranging Eq.~\eqref{eq:basic1} gives
\begin{equation}
\frac{1}{2n}\|X\bm{\Delta}\|_2^2
\le
\frac{1}{n}\bm{\varepsilon}^\top X\bm{\Delta}
+
\lambda\bigl(\|\bm{\beta}\|_1-\|\hat{\bm{\beta}}\|_1\bigr).
\label{eq:basic4}
\end{equation}

\paragraph{Step 2: Dual norm}
For given vectors $\bm a$ and $\bm b$ with the same dimension, we have $|\bm a^T\bm b| \leq \|\bm a\|_\infty \|\bm b\|_1$. Therefore, we obtain
\begin{equation}
\frac{1}{n}\bm{\varepsilon}^\top X\bm{\Delta}
=
\left(\frac{X^\top \bm{\varepsilon}}{n}\right)^\top \bm{\Delta}
\le
\left\|\frac{X^\top \bm{\varepsilon}}{n}\right\|_\infty \|\bm{\Delta}\|_1.
\label{eq:max1}
\end{equation}
Eq. \eqref{eq:max1} corresponds to the step where the stochastic term is bounded by the maximum absolute correlation.

\paragraph{Step 3: Max bound}
Define the event
\begin{equation}
\mathcal{T}
=
\left\{
\left\|\frac{X^\top \bm{\varepsilon}}{n}\right\|_\infty
\le
\frac{\lambda}{2}
\right\}.
\label{eq:event}
\end{equation}
Suppose that $\lambda$ satisfies $\lambda \ge 2\sigma \sqrt{2/n (1+\log \left( \frac{p}{\delta} \right))}$ with $\delta \in (0,1)$. Then we have $P(\mathcal{T}) \ge 1-\delta$ (e.g., \citealp{buhlmann2011statistics}). When $\mathcal{T}$ holds, Eq. \eqref{eq:basic4} implies
\begin{equation}
\frac{1}{2n}\|X\bm{\Delta}\|_2^2 + \lambda \|\hat{\bm{\beta}}\|_1
\le
\frac{\lambda}{2}\|\bm{\Delta}\|_1 + \lambda \|\bm{\beta}\|_1.
\label{eq:max2}
\end{equation}

\paragraph{Step 4: Triangle inequality}
Since $\bm{\beta}_{S^c}=0$, Eq.~\eqref{eq:Delta} implies
\begin{equation}
\bm{\Delta}_{S^c} = \hat{\bm{\beta}}_{S^c},
\qquad
\|\bm{\Delta}\|_1 = \|\bm{\Delta}_S\|_1 + \|\bm{\Delta}_{S^c}\|_1.
\label{eq:decomp1}
\end{equation}
By the triangle inequality,
\begin{equation}
\|\hat{\bm{\beta}}_S\|_1
=
\|\bm{\beta}_S + (\hat{\bm{\beta}}_S-\bm{\beta}_S)\|_1
\ge
\|\bm{\beta}_S\|_1 - \|\hat{\bm{\beta}}_S-\bm{\beta}_S\|_1
=
\|\bm{\beta}_S\|_1 - \|\bm{\Delta}_S\|_1.
\label{eq:triangle}
\end{equation}
Thus, we obtain
\begin{equation}
\|\hat{\bm{\beta}}\|_1
\ge
\|\bm{\beta}_S\|_1 - \|\bm{\Delta}_S\|_1 + \|\hat{\bm{\beta}}_{S^c}\|_1.
\label{eq:triangle2}
\end{equation}
Substituting Eqs.~\eqref{eq:decomp1} and \eqref{eq:triangle2} into Eq. \eqref{eq:max2}, we obtain
%
\begin{equation}
\frac{1}{n}\|X\bm{\Delta}\|_2^2
\le
3\lambda \|\bm{\Delta}_S\|_1-
\lambda \|\bm{\Delta}_{S^c}\|_1.
\label{eq:cone_main}
\end{equation}

\paragraph{Step 5: Drop negative term}

By dropping the nonnegative term $\lambda \|\bm{\Delta}_{S^c}\|_1$ from the right-hand side of Eq.~\eqref{eq:cone_main}, we obtain
\begin{equation}
\frac{1}{n}\|X\bm{\Delta}\|_2^2
\le
3\lambda \|\bm{\Delta}_S\|_1.
\label{eq:before_compat}
\end{equation}

\paragraph{Step 6: Compatibility condition}
Eq.~\eqref{eq:cone_main} implies that $\bm{\Delta}$ satisfies
\[
\|\bm{\Delta}_{S^c}\|_1 \le 3\|\bm{\Delta}_S\|_1.
\]
Applying the compatibility condition in \eqref{eq:compatibility} gives
\begin{equation*}
\|\bm{\Delta}_S\|_1^2
\le
\frac{s\, \bm{\Delta}^\top \hat{\Sigma}\bm{\Delta}}{\phi^2}
=
\frac{s}{\phi^2}
\cdot
\frac{1}{n}\|X\bm{\Delta}\|_2^2.
\label{eq:compatibility4}
\end{equation*}
Taking square roots on both sides and using Eq.~\eqref{eq:before_compat}, we obtain
\begin{equation}
\frac{1}{n}\|X\bm{\Delta}\|_2^2
\le
3\lambda
\sqrt{
\frac{1}{n}\|X\bm{\Delta}\|_2^2
\cdot
\frac{s}{\phi^2}
}.
\label{eq:before_young}
\end{equation}

\paragraph{Step 7: Final step}
Squaring both sides of \eqref{eq:before_young} and rearranging gives the desired upper bound:
\begin{equation*}
\frac{1}{n}\|X\bm{\Delta}\|_2^2
\le
\frac{9\lambda^2 s}{\phi^2}.
\label{eq:final_bound}
\end{equation*}

\end{document}